\documentstyle[prd,aps,epsf,epsfig]{revtex}

\begin{document}
\draft
\twocolumn[\hsize\textwidth\columnwidth\hsize\csname@twocolumnfalse\endcsname
\title{Drifting Spatial Structures in a System with Oppositely Driven Species}
\author{K.-t. Leung${}^{\dagger }$ and R. K. P. Zia${}^{\ddagger }$}
\address{
${}^{\dagger }$Institute of Physics, Academia Sinica,\\
Taipei, Taiwan 11529, ROC\\
${}^{\ddagger }$Center for Stochastic Processes in Science and Engineering\\
Physics Department,\\
Virginia Polytechnic Institute and State University,\\
Blacksburg, Virginia 24061-0435, USA}
\date{\today }
\maketitle

\begin{abstract}
A system consisting of two conservative, 
oppositely driven species of particles
with excluded volume interaction {\em alone\/} is studied on a torus.
The system undergoes a phase transition between a homogeneous
and an inhomogeneous phase, as the particle densities are varied.
Focusing on the inhomogeneous phase with generally 
unequal numbers of the two species,
the spatial structure is found to drift counter-intuitively
against the majority species at a constant velocity that depends
on the external field, system size, and particle densities.
Such dependences are derived
from a coarse-grained continuum theory,
and a microscopic mechanism for the drift is explained.
With virtually no tuning parameter,
various theoretical predictions, notably a field-system-size scaling,
agree extremely well with the simulations.
\end{abstract}

\pacs{PACS Numbers: 64.60.Cn, 05.70.Fh, 66.30.Hs, 82.20.Mj}

]


\section{Introduction}


In the recent decade, there was considerable interest in the statistical
mechanics of a variety of systems in stationary, but {\em non}-equilibrium,
states. Notable examples include fast ionic conductors, surface growth,
electromigration, flux creep in superconductors, propagation of defects and
cracks, electrophoresis, and granular as well as traffic flow. 
Apart from practical applications, the interest lies in the need to
establish a sound foundation for non-equilibrium statistical mechanics, on
par with the Boltzmann/Gibbs formulation for systems in equilibrium. To
pursue these goals, many authors have proposed simple models, just as Lenz
and Ising did\cite{ising} in order to understand the phenomena of phase
transitions of a magnet in thermal equilibrium. Along these lines, Katz,
Lebowitz and Spohn\cite{kls} introduced the simple {\em driven} Ising
lattice gas, as an ``entry'' into the physics of non-equilibrium steady
states. Since then, this field has steadily grown, so that there now exists
many variations and generalizations of the proto model\cite{review}.

One of the most natural generalizations are systems with a second species of
particles. The simplest of these is a model with {\em equal} numbers of
oppositely ``charged'' particles, driven by a uniform external ``electric''
field and diffusing on a periodic, square lattice\cite{shz}. With {\em no}
interparticle interactions, except the excluded volume constraint, this
system exhibits a phase transition, for critical values of the particle
density and external field, from a homogeneous disordered state with sizable
current to an inhomogeneous state with minute current. Particles of the
opposite charge impede each other and ``lock up'' into a dense region. By
symmetry, the average location of this region is time-{\em independent}.
Since its inception, a number of its properties are reasonably well
understood\cite{bsz,fg,vzs,ksz}, while a variety of related ones are proposed%
\cite{bml,ktl,charge-exch}. However, none of these studies focused on a
system with {\em unequal} numbers of the two species. Assuming that a
locked-up state still exists, one should not expect the dense region to
remain stationary. In particular, we can expect a larger number of the
majority species to lie within this region, and so, naturally expect the
block to {\em advance with the majority}. In this paper, we study such
inhomogeneous states with both simulation and analytic techniques. Perhaps
most surprising of the results is that, in the ordered phase, the spatial
structure, as a whole, drifts in a direction {\em opposite }to the intuitive
picture above!

The remainder of this paper is organized as follows. In the next section, we
will provide specifications of this model and some details of the Monte
Carlo runs. We present the simulation results of the counter-intuitive
motion of inhomogeneous states and suggest its microscopic mechanism in
Section III. Section IV will be devoted to the continuum mean-field
approach, which was relatively successful in describing the charge-neutral
model\cite{shz,fg,vzs} and will be re-analyzed for the more general case
here. These theoretical predictions are then compared to the simulation data
in Section V. Particular attention will be paid to the scaling of the drive
with the system size, and the dependence of the drift velocity on control
parameters. We end with some concluding remarks in the last Section.


\section{A Model for Diffusion of Two, Oppositely Biased, Species}


Generalizing the work of Ising, Potts\cite{potts} and Blume, {\em et. al.}%
\cite{beg} introduced models which consists of only three or more states per
site in order to describe various systems such as magnets with spin one or
higher and ternary mixtures. Along similar lines, the natural generalization
of the driven Ising lattice gas\cite{kls} would be models of several species
of particles, driven far from equilibrium by some ``external'' field.
Clearly, there are many physical systems for which such models may be
applicable. Here, we will focus on the simplest one\cite{shz}.

On a square lattice with $L_x\times L_y$ sites, we place $N_{\pm }$
particles with ``charge'' $\pm 1$. At each site, there will be at most one
particle, regardless of its charge. Thus, a configuration of our model is
completely specified by the set of occupation numbers $\left\{ n_{\pm
}\left( x,y\right) \right\} $, where $n_{\pm }\left( x,y\right) =1$ or $0$,
if there is a $\pm $ particle at site $\left( x,y\right) $ or not. Apart
from this excluded volume constraint, there is no interaction between the
particles. However, there is an external ``electric'' field, $E$, chosen to
point in the $+y$ direction, so that a $+$($-$) particle is biased against
moving in the $-$($+$)$y$ direction. Specifically, the system evolves by
random updating. In each trial, a pair of nearest neighbors is randomly
chosen. If it is a particle-hole pair, then the particle hops into the hole
with a probability min$\{ 1, e^{\pm E \hat y \cdot \hat a}\}$ for the $\pm$
species, where $\hat a$ denotes the direction of hopping. $L_x \times L_y$
such trials constitute one time step (or one sweep). Finally, we impose
periodic boundary conditions, so that our lattice is in fact a torus. For
later convenience, let us define the terms ``overall mass density'' and
``overall charge density'', given respectively by

\begin{equation}
m\equiv \frac{N_{+}+N_{-}}{L_xL_y}\quad \text{and}\quad q\equiv \frac{
N_{+}-N_{-}}{L_xL_y}\quad \quad .  \label{m,q}
\end{equation}

Clearly, for $E=0$, there is in fact no distinction between the two species.
The system is purely diffusive and uninteresting. On the other hand, for $%
E>0 $, particles of the opposite charge impede each other. This mutual
blocking is so severe that the system displays drastically different
characteristics if the particle densities are high enough. In all the
previous studies\cite{shz,bsz,fg} of this model, $q$ is restricted to zero
for simplicity, so that there are only two control parameters, $(E,m)$,
besides the system size. There, for fixed $E$, say, the steady state of the
system is disordered and homogeneous, provided $m$ is small enough. By
symmetry, the two opposing particle currents are the same, on the average.
Thus, the (average) hole current, $C$, is zero, while the (average) charge
current, $J$, is non-trivial. As $m$ increases, $J$ increases sublinearly,
as a result of the excluded volume constraint as well as the mutual
blocking. Once $m$ rises beyond a critical value, $m_c(E)$, a phase
transition occurs so that the system is ordered into an inhomogeneous state.
In this state, three regions can be roughly identified: one particle-poor
and two particle-rich zones, one of each species. As might be expected,
these regions span the {\em transverse }dimension of the lattice ($L_x$),
with each particle-rich zone impeding the ``forward motion'' of the other
species. For systems with $O(1)$ aspect ratios, these zones are purely
transverse to the drive, i.e., the densities are {\em homogeneous} in $x$.
The current drops to vanishing values. If $E$ is sufficiently large, this
transition is extremely sharp and dramatic\cite{shz}. With larger aspect
ratios, the system often locks up into somewhat different states, with zones
spanning both $x$ and $y$, i.e., wrapping around the torus with non-trivial
winding numbers\cite{bsz}. The current still suffers a drop, though not to
vanishingly small amounts. In either case, once lock-up occurs, these zones
are stationary on the average, since $C=0$ always.

In this paper, we will study systems with {\em unequal} numbers of the two
species. With $q\neq 0$, many of the previous properties will be different,
although we still expect the presence of a phase transition. For example, in
the homogeneous state, $C$ will not vanish and propagating fluctuations are
possible. Defering a comprehensive study of this model to a later publication%
\cite{promises}, we will focus here only on the {\em inhomogeneous state},
in which the zones are expected to drift. As we will demonstrate and
explain, the system displays a counter-intuitive feature, i.e., the spatial
structures drift in the direction favored by the {\em minority} species. For
example, the inhomogeneities will drift in the negative $y$ direction if $q$
is positive!


\section{Drifting steady-state structures from simulations}


Since our purpose here is to study the novel properties associated with {\em %
unequal} numbers of the two species, we carry out simulations with fixed
particle density ($\bar{\rho}_{+}\equiv N_{+}/L_xL_y$) for the + species and
varying density for the $-$ species ($\bar{\rho}_{-}\equiv N_{-}/L_xL_y<\bar{%
\rho}_{+}$), corresponding to 
\[
q=\bar{\rho}_{+}-\bar{\rho}_{-}>0\quad . 
\]
The behavior for the case of $q<0$ may be deduced simply by symmetry.
Specifically, we choose $\bar{\rho}_{+}=1/4$, $L_x=10,20,40$, $%
L_y=40,160,320 $, and $E$ ranging from about 0.1 to 1. These parameters are
chosen in order to probe the $q>0$ region close to the $q=0$ {\em %
inhomogeneous\/} states near the transition mentioned above, for we expect
the novel properties to be more pronounced there. In contrast, the particles
can hardly move deep in the locked-up phase at higher densities. Simulations
for different $L_x$'s show that the effect of $L_x$ is negligible, as found
for the symmetric, $q=0 $ case\cite{shz}. Thus, unless $L_x\gg L_y$\cite{bsz}
which we will not consider here, we are dealing with a system in which only
one of the dimensions plays an essential role.

Starting from the inhomogeneous state with $\bar\rho_-=\bar \rho_+$, we find
a phase transition into a homogeneous state as we gradually decrease $\bar 
\rho_-$ with $\bar \rho_+$ held fixed. On the $(q,m)$ plane, the phase
boundary between the homogeneous and inhomogeneous states may be located
this way, which is symmetric about the $m$ axis. However, a detailed
discussion of the phase diagram is beyond the scope of this paper.

\begin{figure}[htp]
\epsfig{figure=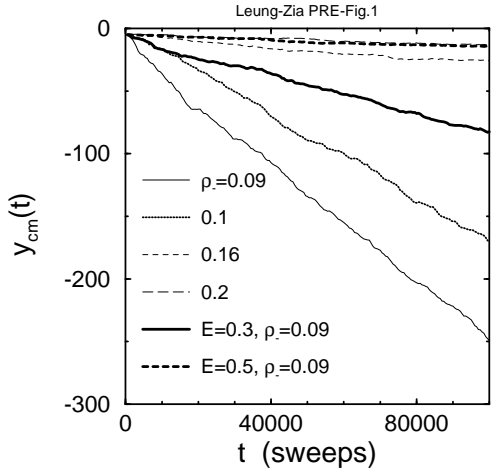,height=3.0in,angle=0} 
\caption{
The position of the center of mass $y_{\rm cm}$ vs time
for the inhomogeneous state.  The steady, backward drift velocity 
decreases with increasing average density of the minority phase.
$\bar\rho_+=0.25$ and $E=0.1976$ except otherwise stated.
}
\label{fig_cmpos}
\end{figure}

\begin{figure}[htp]
\epsfig{figure=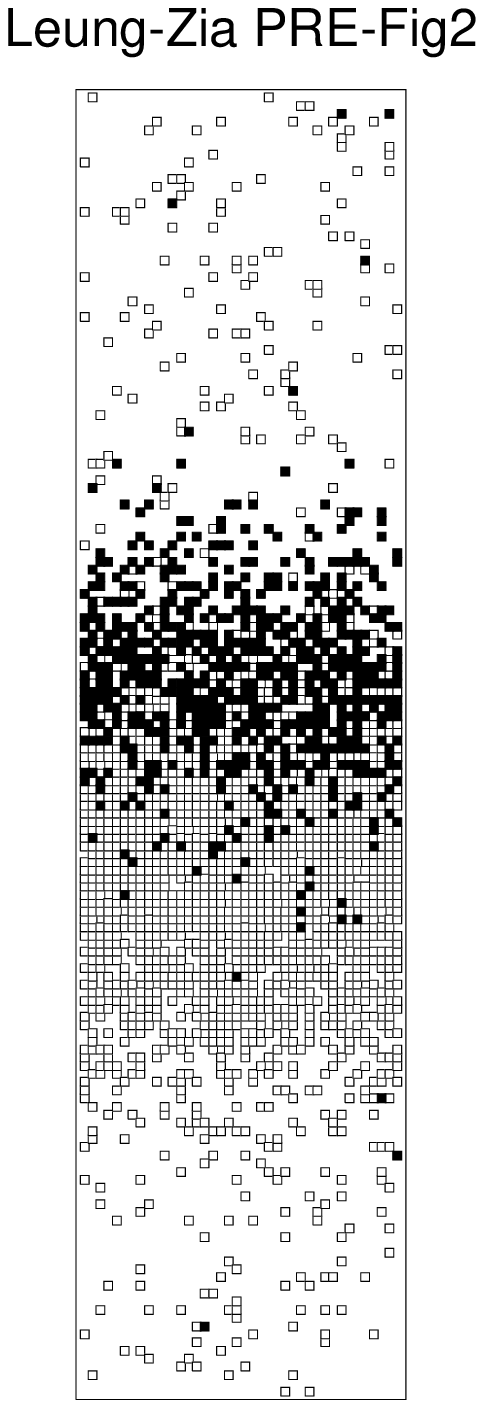,height=3.0in,angle=0} 
\caption{
A typical $40\times 160$ configuration in steady state showing the 
blockage between the two species. The open/filled squares
represent the upward/downward, or $+/-$  drifting species.
Note that there are more + particles escaping through the blockage
to cause the structure to drift {\em downwards\/}.
Here $\bar\rho_+=1/4$, $\bar\rho_-=0.1$, $E=0.1976$.
}
\label{fig_config}
\end{figure}

\begin{figure}[htp]
\epsfig{figure=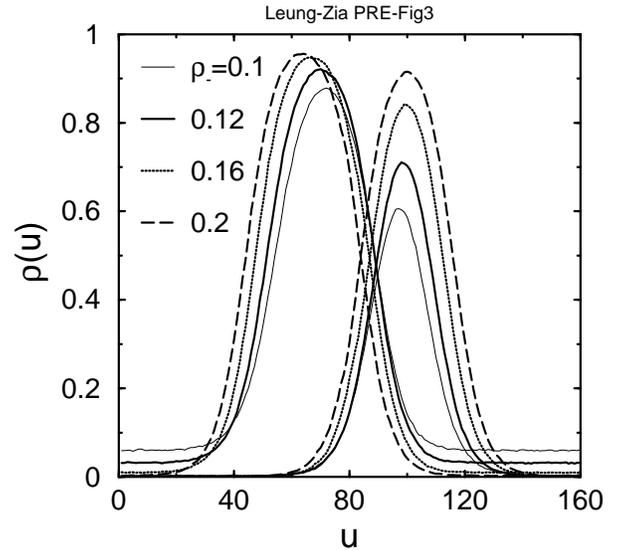,height=3.0in,angle=0} 
\caption{
Steady-state density profiles of a $20\times 160$ system 
for different $\bar \rho_-$, 
at fixed $\bar \rho_+=1/4$ and $E=0.1976$.
$\rho_+(u)$ is on the left, $\rho_-(u)$ on the right, and $u=y-vt$.
}
\label{fig_compares}
\end{figure}

Focusing on the properties of the $q>0$ inhomogeneous states, we find that
the locked-up region of the two species drifts {\em backwards\/} with
respect to the driving direction for the majority (+) species at a definite
velocity $v$ that increases with $q$ but decreases with $E$, as shown in
Fig.~\ref{fig_cmpos}. Fig.~\ref{fig_config} shows a typical inhomogeneous
configuration in steady state for $q>0$. The steady-state ensemble averages
of the local density profiles for the two species, $\rho_{+}(y,t)$ and $%
\rho_{-}(y,t)$, are measured. Due to the drift, they are functions of $%
u\equiv y-vt$ alone. Fig.~\ref{fig_compares} displays the steady-state
density profiles $\rho_\pm(u)$ for various values of $q>0$.

To understand the microscopic mechanism for this backward motion, it is
instructive to consider the role of holes inside the two particle-rich
zones. The probability for a hole to diffuse against the drive into these
zones from the outer zone boundaries is suppressed by $E$ via $e^{-E}$.
Thus, provided $E$ is not too large, there are finite densities of holes
inside the block. For $q=0$, these densities are on the average the same in
the + and $-$ zone. For $q>0$, there are more holes in the $-$ zone because
it is thinner (its thickness given roughly by $\bar{\rho}_{-}L_y$). With
holes available on the inner $(+-)$ zone boundary, a particle may escape
from the block to the particle-poor region through the zone of the {\em %
opposite\/} species. Driven along $E$ then, it eventually returns to the
outer edge of its own zone due to periodic boundary conditions. When a
particle leaves the inner zone boundary, a hole is left behind which may
drift in either direction towards an outer zone boundary, returning to the
hole-rich region. For $q=0$, on the average, the number of holes impinging
on an outer zone boundary equals the number of incoming particles. Thus,
apart from a migration of particles from the inner to the outer zone edges,
the cluster remains stationary. For $q>0$, however, it is relatively easier
for + particles to migrate. It results in more particles than holes
impinging on the outer + zone boundary, and the opposite for the $-$ zone.
It is this imbalance that causes the whole cluster to drift backwards. Of
course it is clear from our argument that $v$ must vanish if $E=\infty $.

In order to see how this arises theoretically and to explore the $E$ and $q$
dependence of $v$, we now turn our attention to a continuum description.


\section{A Continuum Mean-field Description}


Following previous studies of this model\cite{shz,bsz,fg,vzs,ksz}, we rely
on a mean-field type continuum theory to understand the macroscopic
properties here. The equations of motion for the densities, first proposed
in Ref.~\cite{shz}, need no modification and, for completeness, summarized
below (Section III.A). However, the overall constraint on the charge density
will be different, leading to qualitatively new behavior such as drifting
inhomogeneous solutions (Section III.B).

\subsection{Equations of Motion}

To describe the long-wavelength, low-frequency behavior of our model, we
make use of the continuum approach, in which the discrete variables of both
the lattice and the occupation numbers, $n_{\pm }\left( x,y\right) ,$ are
replaced by continuous ones for the densities and space-time: $\rho _{\pm
}\left( {\bf r},t\right) $. For ${\bf r}$, we will continue to write $\left(
x,y\right) $, which should not lead to any confusion, and let $x\in [0,L_x)$%
, etc. The evolution equations of these densities may be ``derived'' by
taking the continuum limits of the mean-field approximation to the Master
equation\cite{ktl}; or they may simply be postulated through considerations
of symmetries. In the former approach, the parameters in the continuum
equation can be related to the microscopic rates. Since we will not be
concerned with the absolute time scale, one parameter may be absorbed into
the definition of $t$. In other words, we will set the diffusion constant,
for the {\em undriven\/} case, to be unity. 
Only one parameter remains, associated with the driving field. If the naive
continuum limit approach is taken, then it is

\begin{equation}
{\cal E}\equiv 2\tanh (E/2)\quad .  \label{E}
\end{equation}

With these considerations, we study the following equations of motion,
written in the form of continuity equations:

\begin{equation}
\frac{\partial \rho _{\pm }}{\partial t}=\nabla \cdot \left[ \phi \nabla
\rho _{\pm }-\rho _{\pm }\nabla \phi \mp \rho _{\pm }\phi {\cal E} {\bf \hat{%
y}}\right] \quad ,  \label{EoM+-}
\end{equation}
where $\phi \equiv 1-\rho _{+}-\rho _{-}$ is the density of holes. Notice
that the first two terms in these equations describe free diffusion of two
distinguishable species of particles. The last term corresponds to the Ohmic
currents, with $\rho _{\pm }\phi $ being the usual density dependent
conductivity. It is also natural to consider the sum and difference of these
equations. Defining $\psi \equiv \rho _{+}-\rho _{-}$ to be the charge
density, they take the form

\begin{eqnarray}
\frac{\partial \phi }{\partial t} &=&\nabla \cdot \left[ \nabla \phi +\phi
\psi {\cal E}{\bf \hat{y}}\right] \quad ,  \label{EoMhole} \\
\frac{\partial \psi }{\partial t} &=&\nabla \cdot \left[ \phi \nabla \psi
-\psi \nabla \phi -\phi (1-\phi ){\cal E}{\bf \hat{y}}\right] \quad .
\label{EoMch}
\end{eqnarray}

These are precisely the equations in Ref.~\cite{shz}. To apply to our
problem, we only need to impose 
\begin{equation}
\frac 1{L_xL_y}\int \psi \ dxdy=q>0  \label{qne0}
\end{equation}
instead of $q=0$. The other constraint, 
\begin{equation}
\frac 1{L_xL_y}\int \phi \ dxdy=1-m\quad ,  \label{mne0}
\end{equation}
as well as the periodic boundary conditions for the densities, of course
remain unchanged.

We may simplify these equations further, by absorbing ${\cal E}$ into the
scale of $y$. There is no need to write new equations, since we can simply
drop ${\cal E}$ from (\ref{EoMhole},\ref{EoMch}) while keeping in mind that $%
L_y$ must be replaced by ${\cal E}L_y$. That the drive provides an intrinsic
length scale implies that the ordinary thermodynamic limit ($L_y\rightarrow
\infty $) must be taken along with ${\cal E}\rightarrow 0$ while holding the
product ${\cal E}L_y$ fixed. However, this simplification may be too
confusing and will not be used here. Due to the central role played by $%
{\cal E}L_y$, let us define 
\begin{equation}
\varepsilon \equiv {\cal E}L_y  \label{EL}
\end{equation}
for future convenience. These equations, (\ref{EoMhole}-\ref{mne0}),
completely specify the dynamics of our model.

\subsection{Inhomogeneous Steady States}

Next, we study steady state solutions to these equations which are spatially 
{\em in}homogeneous. Since we expect the densities to be time dependent, let
us seek solutions with a constant velocity, $v$, namely, $\phi (x,y-vt)$ and 
$\psi (x,y-vt)$. Simplifying further, we note that all the states we
observed in simulations are homogeneous in $x$, so that we will restrict
ourselves to functions of the form:

\begin{equation}
\phi (u)\quad \text{and}\quad \text{ }\psi (u)\quad ,  \label{IH}
\end{equation}
where $u\equiv y-vt.$ Inserting these into (\ref{EoMhole},\ref{EoMch}), we
have 
\begin{eqnarray*}
-v\partial \phi &=&\partial \left[ \partial \phi +\phi \psi {\cal E}\right]
\quad \text{and} \\
-v\partial \psi &=&\partial \left[ \phi \partial \psi -\psi \partial \phi
-\phi (1-\phi ){\cal E}\right] \quad ,
\end{eqnarray*}
where $\partial $ stands for $d/du$. Integrating once, we obtain 
\begin{eqnarray}
\partial \phi +{\cal E}\phi \psi {\cal +}v\phi &=&-C  \nonumber \\
\phi \partial \psi -\psi \partial \phi -{\cal E}\phi (1-\phi ){\cal +}v\psi
&=&-J\quad .  \label{ss-eq}
\end{eqnarray}
The constants, $C$ and $J$, may be interpreted as the two steady state
currents for the holes and the charges respectively in the {\em moving frame}%
. In the ``lab-frame'', the currents should be {\em in}homogeneous, due to
the anticipated drift of the block. As it stands, equations (\ref{ss-eq})
contain three unknown constants ($v,C,J$), which will have to be fixed by
three conditions, i.e., solutions be of period $L_y$, (\ref{qne0}) and (\ref
{mne0}). However, analytically, ($v,C,J$) appear to play more the role of
control parameters while ($L_y,q,m$) will emerge at the end. In this way,
the analytic approach is somewhat opposite to that of simulations, where the
latter (former) are the control (dependent) variables.

To find the solutions, we follow previous studies and introduce variables
which simplify the structure of these equations: 
\begin{equation}
\chi \equiv 1/\phi \quad \text{and\quad }\tilde{\psi}\equiv \psi \chi \quad
.\quad  \label{chi-def}
\end{equation}
Note that, unlike the physical densities which are bounded, $\chi \in \left[
1,\infty \right] $ and $|\tilde{\psi}|\leq \chi -1\in \left[ 0,\infty
\right] .$ Now, (\ref{ss-eq}) becomes 
\begin{eqnarray}
\partial \chi &=&{\cal E}\tilde{\psi}\ +v\chi +C\chi ^2  \label{ODE_hole} \\
\partial \tilde{\psi} &=&{\cal E}(\chi -1)-J\chi ^2-v\tilde{\psi}\chi \quad .
\label{ODE_charge}
\end{eqnarray}
Eliminating $\tilde{\psi}$, we again arrive at an ordinary differential
equation for only one variable: 
\begin{eqnarray}
\chi ^{\prime \prime }-(\chi -1)+\hat{J}\chi ^2 &=&\hat{v}^2\chi ^2+\hat{v}%
\hat{C}\chi ^3  \nonumber \\
&&+\left[ \hat{v}\chi \left( 1-\frac \chi 2\right) +\hat{C}\chi ^2\right]
^{\prime }
\label{ode} 
\end{eqnarray}
where $\hat{J}\equiv J/{\cal E}$, etc. Also, prime denotes $d/du{\cal E}$,
showing again the central role played by ${\cal E}$ in setting the length
scale. For clarity, we have placed on the right hand side of this equation
all the extra terms due to $q\neq 0$.

Unlike in the neutral system, the interpretation of (\ref{ode}) as a
particle ``moving'' in a potential has to be modified, since there are
``velocity'' (i.e., $\chi^\prime$) dependent ``force'' terms. In general,
periodic ``motion'' would be impossible. Of course, here, we must insist on
the existence of such solutions. The consequence is a constraint on the last
term in (\ref{ode}). In particular, multiplying this equation by $\chi
^{\prime }$ and integrating over the full period, we are led to $%
\int(\chi^{\prime})^2\left[\hat{v}\left(1-\chi\right)+2\hat{C}%
\chi\right]du=0.$ Since we are concerned with {\em inhomogeneous} states, we
can expect $(\chi ^{\prime })^2$ to be positive, except for isolated points.
Thus, it is possible to interpret $(\chi ^{\prime })^2du/\int (\chi ^{\prime
})^2 du$ as a new measure on the interval $u\in \left[ 0,L_y\right] $ and
define a new type of average: 
\begin{equation}
\left\langle \bullet \right\rangle \ \equiv \frac{\int \bullet \
(\chi^{\prime })^2du} {\int (\chi ^{\prime })^2du}\quad .  \label{<.>}
\end{equation}
Using this notation, we may write a simple relationship between the drift
velocity and the hole current in the moving frame 
\begin{equation}
v\,\left\langle \chi -1\right\rangle =2C\,\left\langle \chi \right\rangle
\quad .  \label{vC}
\end{equation}
Since $\chi >1$ typically, we conclude that the drift is in the same
direction as the hole current. A similar relationship can be found by
integrating Eq.~(\ref{ODE_hole}), after multiplication by $\phi $. The
result is: 
\begin{equation}
-q{\cal E}\ =v+C\bar{\chi}\quad ,  \label{qvC}
\end{equation}
where bar is the normal average: 
\begin{equation}
\bar{\bullet}\ \equiv \frac{\int \bullet \ du}{L_y}\quad .  \label{bar}
\end{equation}

Eliminating $C$ between (\ref{vC}) and (\ref{qvC}), we see that 
\begin{equation}
\frac v{q{\cal E}}=-\left( 1+\frac{\left\langle \chi -1\right\rangle \bar{%
\chi}}{2\left\langle \chi \right\rangle }\right) ^{-1}  \label{vq}
\end{equation}
is negative definite. So, for example, if the majority species is positive
(i.e., more particles are driven ``upwards''), then the drift of a block
state will be ``downwards''! This behavior is, naively, quite surprising,
since we expect the particle-rich zone to contain more particles of the
majority species so that the entire block should ``advance'' with the
majority. Instead, the block drifts in the {\em opposite} direction. On
closer examination, we find that, since the negative region (in this example
for $q>0$) is thinner, it is easier for positive particles to get through
the blockage. Then, due to the periodic boundary conditions, these particles
pile up ``behind'' the positive region. As a result, the entire block
appears to drift ``backwards.'' This picture simply provides another
perspective on the intutive arguements in Section III. Here, we have proved
that the structure ``retrogrades.''

Clearly, this analysis also leads to $C/q$ being negative, i.e., the holes
moving contrary to the majority species, which is hardly surprising. Before
closing this subsection, we should comment on a number of other constraints
on the three unknowns $(v,C,J)$, independent of the specific values of ($%
L_y,q,m$).

In order to have any periodic solution at all, there must be some form of
restoring ``force'' in (\ref{ode}). Examining the ``potential'' part of this
``force'', i.e., $(\chi -1)-\left( \hat{J}-\hat{v}^2\right) \chi ^2+\hat{v}%
\hat{C}\chi^3$, we see that there would be no ``well'' to trap the particle,
unless 
\begin{equation}
\hat{J}>\hat{v}^2\quad .  \label{constraint 1}
\end{equation}
On the other hand, at least $J-\hat{v}^2<1/4$ is needed, even in the neutral
case. The cubic term further exacerbates the situation. The constraint that
a ``well'' exists turns out to be 
\begin{equation}
\hat{v}\hat{C}\left[ 4-18\left( \hat{J}-\hat{v}^2\right) +27\hat{v}\hat{C}%
\right] <\left( \hat{J}-\hat{v}^2\right) ^2\left[ 1-4\left( \hat{J}-\hat{v}%
^2\right) \right] .  \label{constraint 2}
\end{equation}

More information can be gleaned from regarding (\ref{ODE_hole}) and (\ref
{ODE_charge}) as flows in a ``phase'' plane. If a periodic solution exists,
it would correspond to a closed loop and, by continuity, there would be,
generically, at least one focus (fixed point with spiral orbits) lying
within. In order to have a {\em physical\/} solution, this fixed point must
lie in the physical region: $\chi \in \left[ 1,\infty \right] $ and $|\tilde{%
\psi}|\leq \chi -1.$ Since any fixed point must lie on the curve $\tilde{\psi%
}\ =-\left( \hat{v}\chi +\hat{C}\chi ^2\right) $, we obtain 
\begin{equation}
\left( 1+\hat{v}\right) ^2>-4\hat{C}\quad .  \label{constraint 3}
\end{equation}
Recalling that both $\hat{v}$ and $\hat{C}$ are negative, this confines both
to be small quantities.

In our case, it is easy to check that there are three fixed points, one of
which always lie outside the physical region. Of the remaining two, one is a
focus and the other, a saddle. Based on the characteristics of the neutral
system, we expect our solution curve to run in between these two fixed
points. Unlike the neutral case, however, the flow is not Hamiltonian in
general and, in particular, the focus is not necessarily a center (i.e.,
eigenvalues corresponding to the flow linearized about the fixed point {\em %
not\/} necessarily pure imaginary). Thus, there is no guarantee that we can
find a periodic solution. One possible scenario is that a unique limit cycle
exists for {\em any\/} given $(v,C,J)$, provided they respect the
inequalities (\ref{constraint 1}-\ref{constraint 3}). Another is that $%
(v,C,J)$ satisfy a specific relation which allows for the existence of
periodic solutions. A natural constraint to impose is that this fixed point
be a center, with its associated eigenvalues being {\em pure imaginary}.
This condition is equivalent to setting the coefficient of the last $\chi
^{\prime }$ term in (\ref{ode}) to zero at that fixed point. This gives us
an additional formula for the velocity: 
\begin{equation}
v=\frac{2C}{1-\frac 1{\chi ^{*}}}\quad ,  \label{vfp}
\end{equation}
where $\chi ^{*}$ denotes the fixed-point value of $\chi $ at the center.
Using (\ref{ss-eq}), it is easy to see that the value of the densities at
any fixed point satisfies a cubic equation with parameters $(v,C,J)$.
Eliminating $\chi ^{*}$ by (\ref{vfp}), we then obtain a quintic algebraic
equation for $v$ alone: 
\begin{eqnarray}
8\hat{C}_m^3 &+&32\hat{C}_m^2\hat{v}+42\hat{C}_m\hat{v}^2+2\hat{J}\hat{C}_m%
\hat{v}^2+18\hat{v}^3  \nonumber \\
&+&3\hat{J}\hat{v}^3-\hat{C}_m\hat{v}^4-2\hat{v}^5=0,  \label{center}
\end{eqnarray}
where $C_m\equiv -v-C$ is the mass current in the moving frame. However,
though this scenario guarantees periodic solutions at the lowest order in
the neighborhood of the fixed point, we are unable to prove that, beyond the
linear level, this condition is either necessary or sufficient for the
existence of periodic solutions. There is, nevertheless, some numerical
evidence that such solutions are available, as we will show in the next
Section.Equation (\ref{center}) prescribes a surface $v(C,J)$ in the $C-J$
plane. Subject to numerical uncertainties, we find that the parameters $%
(v,C,J)$ generated by simulations indeed span a surface consistent with this
scenario. 


\section{Field-Size Scaling and Comparisons with Simulations}

\begin{figure}[htp]
\epsfig{figure=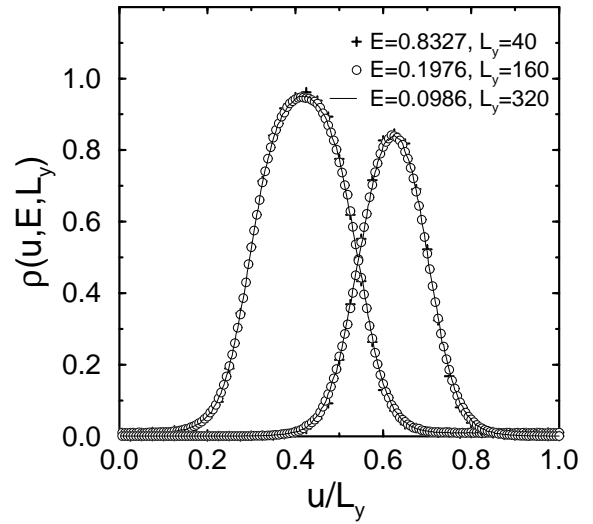,height=3.0in,angle=0} 
\caption{
Predicted field-size scaling is confirmed by simulations
for different $E$ and $L_y$ with fixed ${\cal E}L_y$.
$L_x=20$, $\bar\rho_+=1/4$, $\bar\rho_-=0.16$.
}
\label{fig_fss}
\end{figure}

To determine how close the continuum model corresponds to the discrete
model, we subject our theoretical predictions to the tests of simulations.
The first is concerned with the scaling behavior in the system size and
field strength. Choosing the alternative set of control parameters $(L_y,q,m)
$ in favor of $(v,C,J)$, Eqs.~(\ref{ODE_hole}) and (\ref{ODE_charge})
implies that the solutions for the densities obey a simple scaling form (cf. 
\cite{shz,fg,ktl}): 
\begin{eqnarray}
\chi (u,E,L_y,q,m) &=&\tilde{F}_\chi ({\cal E}u,{\cal E}L_y,q,m)  \nonumber
\\
&=&F_\chi (u/L_y,{\cal E}L_y,q,m),
\end{eqnarray}
where $\tilde{F}_\chi $ and $F_\chi $ are appropriate scaling functions.
Similar scaling form holds for $\psi $, of course. Monte Carlo simulations,
with fixed $\bar{\rho}_{+}$ and varying $\bar{\rho}_{-}$, using a wide range
of $L_y$ and $E$ with {\em fixed\/} $\varepsilon ={\cal E}L_y$ show
excellent agreement. An example is shown in Fig.~\ref{fig_fss}. One
immediate implication of this result is that the thermodynamic limit has to
be taken with care, as phase transitions survive only in the double limits $%
E\to 0$ and $L_y\to \infty $ with $\varepsilon $ held fixed.

\begin{figure}[htp]
\epsfig{figure=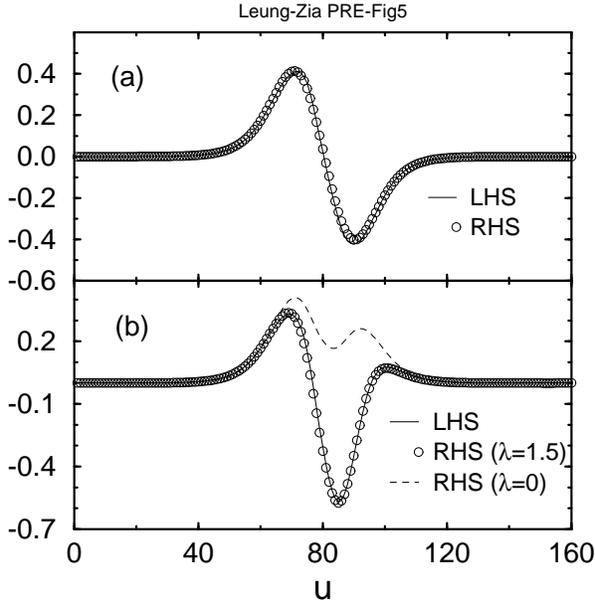,height=3.0in,angle=0} 
\caption{
Typical tests of local properties
of the continuum model against simulations for
(a) the hole equation 
(12),
and (b) the charge equation 
(13),
demonstrating the significance
of the correction term $\propto \lambda {\cal E}\rho_+\rho_-\chi$
in the latter.
$L_x=40$, $L_y=160$, $E=0.1976$, $\bar\rho_+=1/4$, and $\bar\rho_-=0.1$.
}
\label{fig_eqn}
\end{figure}

A more stringent test is to check to what extent the data actually satisfy
the differential equations (\ref{ODE_hole}) and (\ref{ODE_charge}). In the
continuum description, it is more natural to use the currents in the moving
frame because they are the integration constants. In simulations, the
(spatial and temporal) average currents in the lab frame are more
accessible. They are related. For example, the hole current 
in the lab frame is given by $C+v\bar{\phi}$, which is greater than $C$, in
magnitude. With {\em no tuning parameter\/}, the equation (\ref{ODE_hole})
for the hole density fits the data very well (see Fig.~\ref{fig_eqn}(a)),
but there are appreciable discrepancies in the equation (\ref{ODE_charge})
for the charge density in the particle-rich region (see dashed line in Fig.~%
\ref{fig_eqn}(b)). Similar discrepancies were also observed in a closely
related model consisting of two species driven along orthogonal directions%
\cite{ktl}. In that model, the asymmetry between the $(+-)$ and $(-+)$
nearest-neighbor correlations along the field direction was shown to give
rise to additional cubic terms of the form $\pm \lambda \rho _{+}\rho
_{-}\phi {\cal E}\hat{y}$ in the currents for the $\pm $ species, which
enter inside the brackets on the RHS of Eq.~(\ref{EoM+-}). Due to the
opposite signs, they cancel out in the equation (\ref{EoMhole}) for the hole
(or mass) density but contribute an extra term to equation (\ref{EoMch}) for
the charge density. These terms represent the lowest order corrections to
our mean-field equations in Section IV. After such a term $-2\lambda {\cal E}%
\rho _{+}\rho _{-}\chi $ is added to the RHS of Eq.~(\ref{ODE_charge}),
significant improvements are found, for a suitable proportional constant $%
\lambda $ (solid line in Fig.~\ref{fig_eqn}(b)).

\begin{figure}[htp]
\epsfig{figure=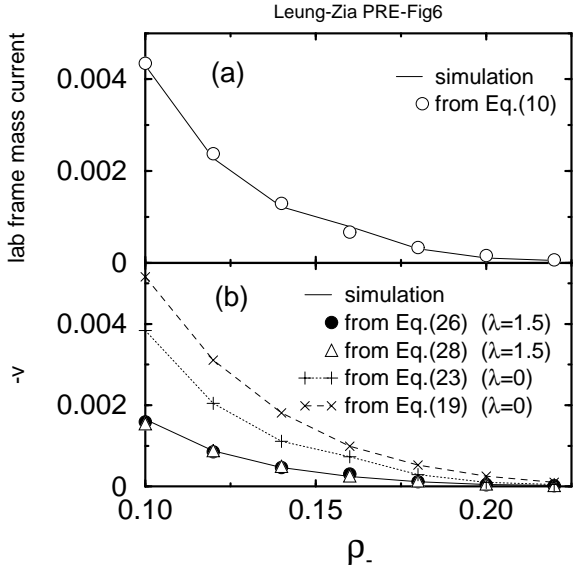,height=3.0in,angle=0} 
\caption{
Comparison between theory and simulation for
(a) the mass current in the lab frame, $-C-v\bar\phi$, and 
(b) the (negative) drift velocity with and without the correction term,
as a function of the average density of the minority species.
$L_x=20$, $L_y=160$, $E=0.1976$, and $\bar\rho_+=1/4$.
}
\label{fig_v}
\end{figure}

Further comparisons are concerned with the mean current and drift velocity.
The mass current $-C-v\bar{\phi}$ in the lab frame, finite for $q>0$, is
simply given by $\overline{\phi \psi }{\cal E}$, which is obtained by
integrating the first of Eq.~(\ref{ss-eq}). Excellent agreement with
simulations is found, as shown in Fig.~\ref{fig_v}(a). This comparison does
not involve the $\lambda $ correction terms. Other predictions, such as (\ref
{vq}) and (\ref{vfp}), derived by using both the hole and charge equations
without $\lambda $, do not agree as well. Fig.~\ref{fig_v}(b) exhibits
increasing deviations as $q$ increases. The agreement, however, can again be
significantly improved by including the $\lambda $ terms, with the same
choice of $\lambda \approx 1.5$ as in Fig.~\ref{fig_eqn}. With $\lambda \neq
0$, (\ref{vq}) is slightly modified: 
\begin{equation}
\frac v{q{\cal E}}=-\left( 1+\frac{2\lambda +\left\langle \chi
-1\right\rangle _\lambda \bar{\chi}}{2(1-\lambda )\left\langle \chi
\right\rangle _\lambda }\right) ^{-1}\quad ,  \label{vq_lambda}
\end{equation}
where, similar to (\ref{<.>}), 
\begin{equation}
\left\langle \bullet \right\rangle _\lambda \ \equiv \frac{\int \bullet \
(\chi ^{\prime }/\chi ^\lambda )^2du}{\int (\chi ^{\prime }/\chi ^\lambda
)^2du}\quad .  \label{<.>lambda}
\end{equation}
Also, (\ref{vfp}) becomes 
\begin{equation}
v=\frac{(2-\lambda )C}{1+\frac{\lambda -1}{\chi ^{*}}}\quad ,
\label{vfplambda}
\end{equation}
where $\chi ^{*}$ is approximated by the spatial maximum $\chi _{{\rm max}}$
in Fig.~\ref{fig_v}, the error incurred is very small as both quantities are
much greater than 1. Detailed discussions and derivations of the $\lambda $
terms will be presented elsewhere\cite{promises}.

\begin{figure}[htp]
\epsfig{figure=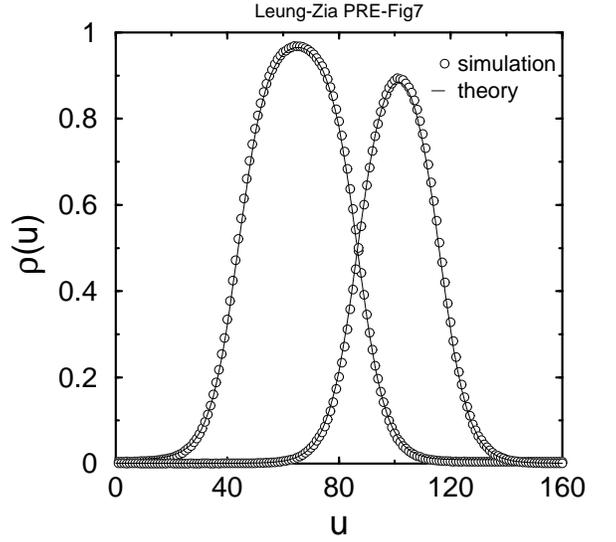,height=3.0in,angle=0} 
\caption{
Excellent agreement between theory and simulation for the density profiles.
$L_y=160$, $E=0.1976$, $\bar\rho_+=0.272$, $\bar\rho_-=0.184$ and $\lambda=0$.
}
\label{fig_compare}
\end{figure}

The final convincing evidence for the quantitative agreement is a direct
comparison of the density profiles. For simplicity, we consider only the
case of $\lambda=0$. The fixed-point condition mentioned near the end of
Sec.~IV picks out a unique $v$ for a given set of currents $(J,C)$ via (\ref
{center}). Equations (\ref{ss-eq}) then contain no free parameter, and we
can obtain the profiles by numerical integration, using for instance the
Runge-Kutta method\cite{recipes}. A typical comparison with the simulated
profiles using the same set of parameters $(q,m,L_y,E)$ is presented in
Fig.~\ref{fig_compare}. The agreement for this case of rather small
$q\approx 0.09$ is again excellent, even without $\lambda$. 
We expect more deviations for larger $q$, where the $\lambda$ terms
can no longer be ignored.

These comparisons provide strong support to our claim that the continuum
model, which may be systematically refined if necessary, represents a
surprisingly accurate description of the simulated discrete model.


\section{Concluding Remarks}


To summarize, we have studied, using both Monte Carlo techniques and the
continuum meanfield method, a diffusive system of two species of particles,
driven in opposite directions by an external field. With {\em periodic}
boundary conditions imposed, this system settles into a {\em non-equilibrium}
state with a steady current. For simplicity, we have restricted ourselves to
non-interacting particles, apart from an excluded volume contraint. Thus, as
the particle densities increase, this system undergoes a phase transition,
from a homogeneous disordered phase with a high current to an inhomogeneous
one with minute current. In the latter state, the two species impede each
other so much that they form a blockage of high, local particle density.
When the particle densities are {\em unequal}, this spatial structure
displays a counter-intuitive behavior. It drifts with a constant velocity,
in a direction {\em opposite} to that favored by the majority species.
Remarkably, simulation results agree reasonably well with most aspects of
the theory, especially the prediction of ${\cal E}L_y$ scaling.

On the other hand, there are clear signs of disagreement, mainly in
connection with (\ref{ODE_charge}). Since our theory is based on meanfield
assumptions, one avenue for improvement is to take some correlations into
account. The simplest addition, involving cubic terms\cite{ktl} in (\ref
{EoM+-}) lead to significant improvements. Encouraged by these findings, we
are undertaking a comprehensive study, including a general phase diagram in
the $(q,m)$ plane, of the effects of such terms. In this paper, we have
focused only on the drifting inhomogeneous state. Although we have performed
some analysis for the {\em homogeneous} state \cite{promises}, much remains
to be investigated. For example, it would be desirable to observe, in
simulations, the drift of fluctuations from the uniform densities. Of
course, as in the neutral case, we should expect long range correlations\cite
{ksz}.

When restricted to one-dimension (i.e., one column), this model is exactly
soluable\cite{derrida,leeuwen}, since the order of any particular string of
+'s and $-$'s is invariant and no phase transitions can occur. With open
boundary conditions, it can be mapped onto the Rubinstein-Duke model for
electrophoresis\cite{rub-duke}, and more interesting phenomena can occur.
Beyond the simple model studied here, there are many other generalizations
which may be relevant to a variety of physical systems. We mention only a
few here.

The existence of the inhomogeneous state depends crucially on the mutual
blocking between the species. To find out the importance of this effect, we
may introduce ``charge exchange'' processes which takes place at a fraction
of the particle-hole exchange rate. As in the neutral case\cite{charge-exch}%
, we can expect to find the transition between the disordered and the
inhomogeneous states to be both continuous and discontinuous. It would be
interesting to map out a complete phase diagram in the $({\cal E}L_y,m,q)$
space. Further, such a system can display interesting behavior even in one
dimension\cite{1-d}, especially with open boundaries\cite{1-dob}. In
particular, it is closely related to the model of ``first and second class
particles'' (e.g., cars and trucks) moving on a ring, in which a first class
particle is allowed to overtake a second class one with some rate\cite
{car-truck}. In that case, there are distinct phases, with properties
reminiscent of our inhomogeneous states. The qualitative vs. quantitative
similarities should be investigated.

Another generalization involves the two species being driven in orthogonal
directions. These models are motivated by the phenomena of traffic flow in
city blocks and display a considerable variety of phases\cite{bml,ktl}.
However, we believe that there are no studies with {\em unequal} numbers of
the two species (though we are aware of a study with varying densities of a 
{\em third} species\cite{others}). As we have shown in this paper, it is
likely that novel behavior will be found if the species are not exactly
balanced, which in view of the physical motivations should be the more
generic cases to study.

In all the models mentioned so far, there is {\em no} interparticle
interaction, except for the excluded volume constraint. It is clearly
important to ask what the effects of including such interactions are. In
particular, even in the absence of external drives, there would be rich
phase diagrams if interactions were present\cite{beg,potts}. Thus, it is
natural to inquire how the drive would modify these phase transitions. We
are aware of only one study of a driven system with two interacting species%
\cite{plg}. Though the regime investigated was extremely limited, several
novel features were found.

Finally, we point out that, in physical systems such as fast ionic
conductors, the two species may be of different mobilities and different
``charges''. These properties add two entirely new dimensions to the phase
space, leading to seemingly endless horizons for future explorations.


\vspace{0.4cm} 
\center{\bf ACKNOWLEDGEMENTS}


We thank B. Schmittmann, S. Sandow and S. Cornell for many stimulating
discussions. One of us (RKPZ) is grateful for the hospitality of Academia
Sinica (Taipei), where some of this work was carried out. This research is
supported in part by grants from the National Science Council and National
High Performance Computing Center of ROC, and the US National Science
Foundation through the Division of Materials Research.



\vspace{1cm}



\end{document}